\documentclass[%
 twocolumn,
 notitlepage,
superscriptaddress,
showpacs,preprintnumbers,
amsmath,amssymb,
aps,
prx,
floatfix,
 ]{revtex4-2}
\usepackage{physics}
\usepackage{graphicx}
\usepackage{dcolumn}
\usepackage[mathscr]{euscript}
\usepackage{ragged2e}
\usepackage{bm}
\usepackage[shortlabels]{enumitem}
\usepackage[margin=.75in]{geometry}
\usepackage{amsmath}
\usepackage{amssymb}
\usepackage{float}
\usepackage{color}
\usepackage{hyperref}

\begin{document}

\title{Enhanced robustness and dimensional crossover of superradiance in cuboidal nanocrystal superlattices}
\author{Sushrut Ghonge}
\email{sghonge@nd.edu}
\affiliation{Department of Physics and Astronomy, University of Notre Dame, Notre Dame, Indiana 46556, USA}

\author{David Engel}
\affiliation{Department of Physics and Astronomy, University of Notre Dame, Notre Dame, Indiana 46556, USA}
 
\author{Francesco Mattiotti}
\affiliation{University of Strasbourg and CNRS, CESQ and ISIS (UMR 7006), aQCess, 67000 Strasbourg, France}
 
\author{G. Luca Celardo}
\affiliation{Department of Physics and Astronomy, University of Florence, INFN, Florence Section, and CSDC, Via Sansone 1, 50019 Sesto Fiorentino, Firenze, Italy}

\author{Masaru Kuno}
\affiliation{Department of Physics and Astronomy, University of Notre Dame, Notre Dame, Indiana 46556, USA}
\affiliation{Department of Chemistry and Biochemistry, University of Notre Dame, Notre Dame, Indiana 46556, USA}

\author{Boldizs\'{a}r Jank\'{o}}
\email{bjanko@nd.edu}
\affiliation{Department of Physics and Astronomy, University of Notre Dame, Notre Dame, Indiana 46556, USA}

\begin{abstract} 
Cooperative emission of coherent radiation from multiple emitters (known as superradiance) has been predicted and observed in various physical systems, most recently in CsPbBr$_3$ nanocrystal superlattices. Superradiant emission is coherent and occurs on timescales faster than the emission from isolated nanocrystals. Theory predicts cooperative emission being faster by a factor of up to the number of nanocrystals ($N$). However, superradiance is strongly suppressed due to the presence of energetic disorder, stemming from nanocrystal size variations and thermal decoherence. Here, we analyze superradiance from superlattices of different dimensionalities (one-, two- and three-dimensional) with variable nanocrystal aspect ratios. We predict as much as a 15-fold enhancement in robustness against realistic values of energetic disorder in three-dimensional (3D) superlattices composed of cuboid-shaped, as opposed to cube-shaped, nanocrystals. Superradiance from small $(N\lesssim 10^3)$ two-dimensional (2D) superlattices is up to ten times more robust to static disorder and up to twice as robust to thermal decoherence than 3D superlattices with the same $N$. As the number of $N$ increases, a crossover in the robustness of superradiance occurs from 2D to 3D superlattices. For large $N\ (> 10^3)$, the robustness in 3D superlattices increases with $N$, showing cooperative robustness to disorder. This opens the possibility of observing superradiance even at room temperature in large 3D superlattices, if nanocrystal size fluctuations can be kept small. 
\end{abstract}
\maketitle

\section{Introduction}
In 1954, Dicke \cite{dicke1954coherence} predicted a phenomenon he termed superradiance (SR), which involves a collection of identical light emitters spontaneously emitting intense coherent radiation. Since then, superradiance has been observed in various physical systems such as molecular aggregates \cite{de1990dephasing,fidder1990superradiant}, cold atoms \cite{araujo2016superradiance}, diamond nanocrystals \cite{bradac2017room}, semiconductor quantum dot ensembles \cite{brandes2005coherent,scheibner2007superradiance}, and more recently in both nanocrystal (NC) superlattices \cite{raino2018superfluorescence,philbin2021room,cherniukh2021perovskite} and hybrid perovskite thin films \cite{findik2021high}. SR can be used to develop ultra-narrow linewidth lasers for quantum metrology \cite{bohnet2012steady}, efficient light-harvesting and photon detecting devices \cite{higgins2014superabsorption} and is key to recently proposed sun-light pumped lasers \cite{biolaser2022}.

Typically, the probability density that a single excited particle emits a photon is exponentially distributed and is characterized by a lifetime $\tau$. If there are $N$ entangled emitters, all in the excited state, SR theory predicts cooperative emission with $\sim N^2$ times higher peak intensity than that of a single emitter (in contrast to $\sim N$ for unentangled emitters). If the incident radiation is so weak that only one excitation is present, single excitation superradiance \cite{scully2009super} can result. In this case, if the excitation is coherently shared by $N$ emitters, its lifetime decreases by a factor of $N$ and, in turn, causes the emission intensity to scale by the same factor.

Experimentally, superfluorescence (a type of superradiance where emitting dipoles are spontaneously entangled) has recently been observed in three-dimensional (3D) superlattices of $10^6 - 10^8$ cube-shaped CsPbBr$_3$ NCs \cite{raino2018superfluorescence}. At low fluence, the observed radiative decay rate is only a factor of three faster than that of individual nanocrystals. This is several orders of magnitude smaller than what is predicted from a linear $N$ scaling. We have recently shown that this deviation can be rationalized \cite{mattiotti2020thermal}. 

We use the radiative Hamiltonian approach to model SR from NC superlattices, which is valid for all systems sizes in the point-dipole approximation. This is in contrast to static dipole-dipole coupling, used by for example, Blach \textit{et al}. \cite{blach2022superradiance}, which is only valid for distances smaller than the emission wavelength. The radiative Hamiltonian was first used to study lattices of atomic dipoles \cite{bettles2015cooperative,bettles2017cooperative,bellando2014cooperative} and then successfully applied to study superradiance from perovskite NC superlattices \cite{mattiotti2020thermal} by some of the authors of the present paper. Sierra \textit{et al}. \cite{sierra2022dicke} also use a similar Hamiltonian to study the effect of dimensionality on the critical distance needed to yield superradiance in arrays of atoms that have a single excited state. Whereas we model isotropic emission from NCs by considering three excited states along $x, y,$ and $z$ directions.

Through detailed modeling, we find that both energetic disorder, stemming from NC size and band gap variations, as well as the collective effects of thermal decoherence (i.e., thermal noise due to a finite temperature) suppress SR rate enhancements. Thus while superradiant enhancements of up to $O(10^3)$ are possible with a few thousand coupled NCs in a superlattice, thermal decoherence and static disorder bring this enhancement factor down to $O(1)$.

Having established the origins of a smaller than expected superradiant enhancement, we discuss approaches to overcoming this suppression. SR occurs when nanocrystals couple with each other through interactions between their transition dipoles. Dipole-dipole interaction strengths are inversely proportional to the cube of their separation. Bringing dipoles closer therefore increases the coupling and should make superradiance more robust to disorder. The interaction also depends on the relative orientations of the dipoles as well as the angles between them and the mutual vector joining them. If NCs are all arranged in a line (one dimension, 1D) or a plane (two dimensions, 2D), all vectors joining dipoles lie on the same line or in the same plane, respectively. This changes the structure of dipolar couplings and affects how systems respond to static and thermal disorder.

In this study we establish how changing superlattice dimensionality affects superradiance and its robustness to static and thermal disorder. We also investigate the effect changing component NC aspect ratios has on SR and its robustness. This is done by simulating SR from cuboidal NC superlattices. Our results show an interplay between superlattice dimensionality and NC shape on SR enhancement. We show that superlattices of cuboidal NCs are more superradiant and more robust to static and thermal disorder for any number of NCs. Moreover, SR from 2D superlattices can be more robust to disorder if they are composed of a small number of NCs ($N < 10^3$). On the other hand, for large superlattices, robustness to disorder increases with superradiant decay rate. This effect has been predicted previously and is known as ``cooperative robustness'' \cite{celardo2014cooperative,celardo2014cooperative2,chavez2019real}. Here we demonstrate cooperative robustness in a realistic model of perovskite superlattices. Since 3D superlattices show larger increases of the superradiant decay rate with system size, they possess cooperative robustness that can yield superradiance even in the presence of disorder having energies comparable to room-temperature thermal energy.

\begin{figure*}[htpb]
    \includegraphics[width=0.8\linewidth]{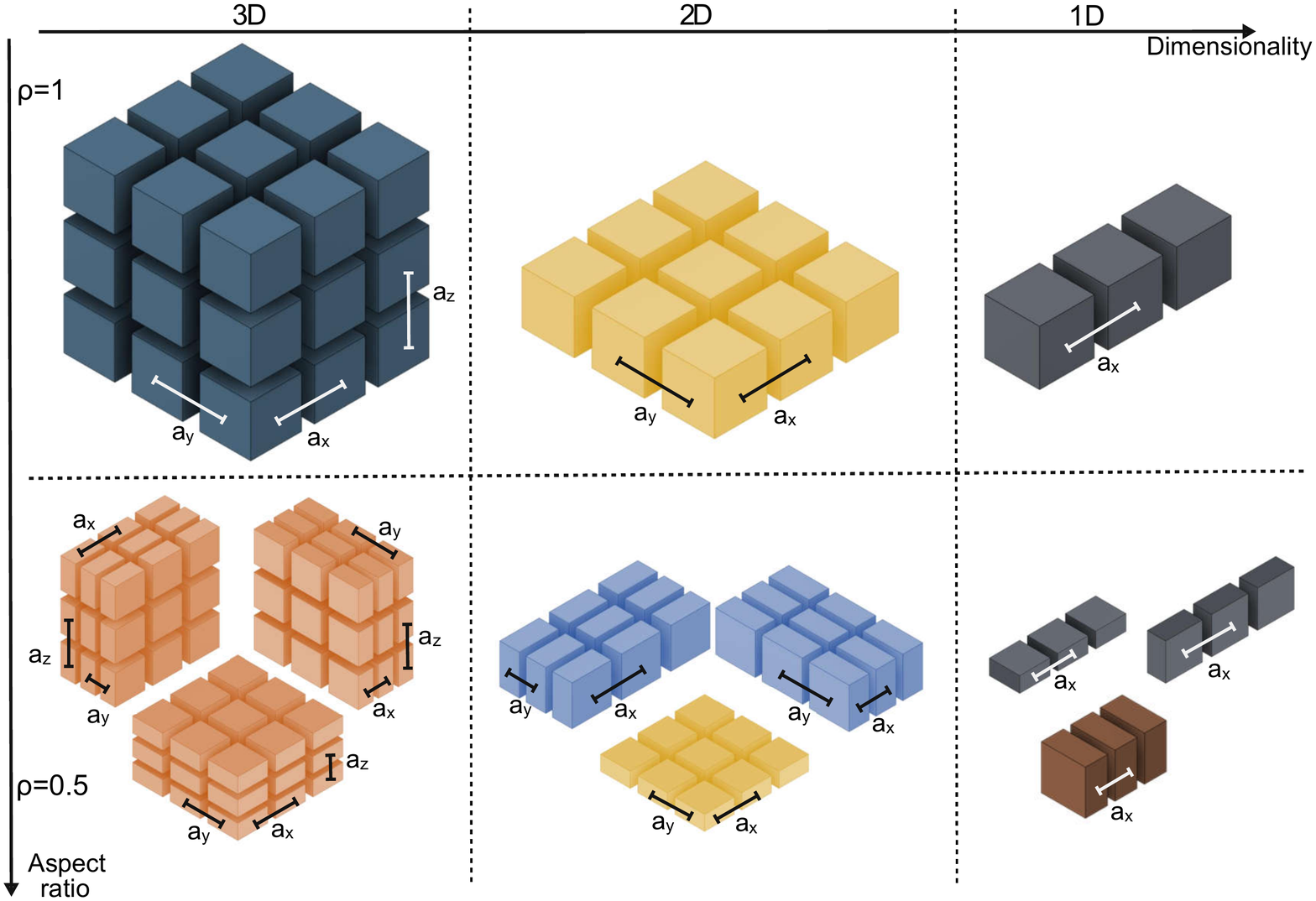}
    \caption{Schematic Figs. showing superlattices with different dimensionalities and center-to-center distances. The top row shows superlattices of cubic NCs while the bottom row shows superlattices of cuboidal NCs. The key parameters in our model, the center-to-center distances along different axes, are labeled on all Figs.. In 3D, the three different ways to arrange cuboidal NCs into a superlattice are all equivalent (orange Figs.). In 2D, there is one way to arrange cuboidal NCs (yellow) that does not change the center-to-center distances and two equivalent ways (blue) that do. In 1D there is only one way (brown Fig.) to arrange cuboidal NCs that modifies the center-to-center distance.}
    \label{fig:sls}
    \hrulefill
\end{figure*}

\section{Theoretical Model and Numerical Methods}

A superlattice is defined here as a lattice consisting of $N_x, N_y, \text{ and } N_z$ NCs self-assembled along $x, y, \text{ and } z$ directions, respectively. Corresponding NC center-to-center distances are $a_x, a_y, \text{ and } a_z$, which are sums of NC edge lengths and interparticle spacings along a given direction. The total number of NCs ($N$) in a superlattice is $N=N_x\times N_y \times N_z$.

When $N_x= N_y= N_z > 1$, the arrangement is said to be a regular 3D superlattice. If the number of NCs along any one edge (say $N_z$) is equal to 1, and $N_x=N_y>1$, the arrangement is a regular 2D superlattice. In a 1D superlattice, NCs are arranged only along one direction (say, $N_x=N_y=1 < N_z$). \textbf{Fig. \ref{fig:sls}} illustrates superlattices of different dimensionalities.

We define superlattice aspect ratio, $\rho$, as the ratio of center-to-center distance along shorter and longer superlattice axes. In 3D we assume that center-to-center distances along $x$ and $y$ axes (e.g., $a_x$ and $a_y$) are equal so that $\rho\equiv a_z/a_x$. In 2D, there are only two axes (e.g., $x$ and $y$). Consequently, $\rho = a_x/a_y$. In 1D, aspect ratio is not defined because there is only one axis. \textbf{Fig. \ref{fig:sls}} shows the various ways in which one can arrange cuboidal NCs into superlattices with different $\rho$ values.

SR from NC superlattices is modeled via lattices of interacting transition dipoles. Such dipole lattices have previously been studied theoretically and reveal the emergence of cooperative behavior when lattices are free of disorder \cite{bettles2015cooperative,bettles2017cooperative,bellando2014cooperative}. In these models, each lattice site has a point dipole that can be oriented in one of three directions ($x,\ y,\text{ or }z$). Point dipoles interact via their mutual radiation field. In this way, SR from cold atomic clouds has been predicted and modeled \cite{bienaime2012controlled,bienaime2013cooperativity,rouabah2014coherence}.

Modeling is done in the so-called `single-excitation superradiance' regime to capture behavior induced by low excitation intensities employed in actual measurements \cite{roof2016observation,tighineanu2016single}. In this limit, there is only one excited NC in the superlattice. This simplification reveals relevant SR trends because the involved Hilbert space scales as $N$, in contrast to the $2^N$ scaling in the high fluence limit. The approach has previously allowed us to numerically study large superlattices of $\sim 10^4$ interacting dipoles \cite{mattiotti2020thermal}.

NCs with edge lengths of $l=9-29$ nm are studied \cite{brennan2017origin,toso2021metamorphoses,philbin2021room,sui2021zone}. This corresponds to center-to-center distances of $10-30$ nm, assuming an interparticle spacing of 1 nm. Such NCs lie in the intermediate to weak confinement regime given a $\mathrm{CsPbBr_3}$ Bohr exciton radius ($a_B$) of 3.5 nm \cite{becker2018bright,brennan2017origin}. For NC dimensions above $a_B$, band gaps do not vary significantly \cite{becker2018bright}, allowing an assumption that transition dipole moments are size independent \cite{mattiotti2020thermal}. The spacing between NCs is due to a coating of organic ligands, specifically oleic acid and oleylamine \cite{raino2018superfluorescence,baranov2019investigation,baranov2020aging}. The optical dielectric constants of CsPbBr$_3$ \cite{becker2018bright}, oleic acid \cite{o2006merck}, and oleylamine \cite{horikoshi2011microwave} are 4.8, 3.1, and 2.13, respectively. The exciton structure and coupling can be affected by this dielectric mismatch when the size of the NCs is smaller that the exciton Bohr radius \cite{movilla2020dielectric}, thus being another reason for us to only consider larger NCs.

We use following non-Hermitian Hamiltonian \cite{mattiotti2020thermal} to model the NC superlattice,

\begin{align}\label{hamiltonian}
    \hat{H} = & \hbar\Big(\omega_0 - i\frac{\gamma_r}{2} \Big)\sum_{\alpha}\sum_{n=1}^{N} \ket{n,\alpha}\bra{n,\alpha}\\ 
    & + \sum_{\alpha,\beta}\sum_{\substack{m,n=1 \\ m\neq n}}^{N} J_{mn}^{\alpha\beta} \ket{m,\alpha}\bra{n,\beta}, \nonumber
\end{align}

where $\hbar \omega_0 (\equiv E_0)$ is the NC band gap, $\gamma_r$ is the single NC radiative decay rate, and $\alpha,\beta=x,y,z$ are transition dipole directions. $J_{mn}^{\alpha\beta}$ terms, which indicate the coupling between NCs, are given by
\begin{align}\label{realoffdiagonal}
    \text{Re}(J_{mn}^{\alpha\beta}) = & \frac{\hbar\gamma_r}{2}\Big[ y_0 (kr_{mn}) \hat{e}_\alpha.\hat{e}_\beta - \frac{1}{2}y_2(kr_{mn})g_{mn}^{\alpha\beta}
    \Big],
\end{align}
\begin{align}\label{imagoffdiagonal}
    \text{Im}(J_{mn}^{\alpha\beta}) = - & \frac{\hbar\gamma_r}{2}\Big[ j_0 (kr_{mn}) \hat{e}_\alpha.\hat{e}_\beta - \frac{1}{2}j_2(kr_{mn})g_{mn}^{\alpha\beta} \Big].
\end{align}

Here, $g_{mn}^{\alpha\beta}$ denotes the geometric factor $[\hat{e}_\alpha.\hat{e}_\beta - 3(\hat{e}_\alpha.\hat{r}_{mn})(\hat{r}_{mn}.\hat{e}_\beta)]$, and $y_\nu$ and $j_\nu$ are spherical Bessel functions of order $\nu$. $\vec{r}_{mn}$ is the vector joining the centers of the $m$-th and $n$-th NCs ($r_{mn}$ and $\hat{r}_{mn}$ are the magnitude of $\vec{r}_{mn}$ and the unit vector along $\vec{r}_{mn}$ respectively), and $k=\frac{2\pi}{\lambda}$ is the radiation wavenumber inside the material. The vectors $\vec{r}_{mn}$ depend on the superlattice parameters, $a_x, a_y$ and $a_z$ that we defined above. In our model, when we change the NC sizes and aspect ratios, the corresponding center-to-center distances $\vec{r}_{mn}$ will also change. CsPbBr$_3$ parameters of $E_0=2.38$ eV and $\lambda=237$ nm are used, corresponding to a vacuum wavelength of 520 nm \cite{raino2018superfluorescence} for a refractive index of 2.2 \cite{becker2018bright}. Note that the model is general and applicable to any NC superlattice provided appropriate material parameters.

Equation~\eqref{realoffdiagonal} is a generalized dipole-dipole coupling term. For small distances ($kr_{mn}\ll 1$), such as when we consider the nearest neighbor coupling, it is the usual dipole-dipole interaction that is proportional to $1/r_{mn}^3$. The imaginary term, Equation~\eqref{imagoffdiagonal}, arises due to the field-mediated interaction between dipoles. We define the nearest neighbor coupling ($\mathcal{J}$) as the maximum coupling between transition dipoles of neighboring NCs, $\mathcal{J} = \max\limits_{\alpha,\beta}\{\abs{J_{m,m+1}^{\alpha\beta}}\}$. The real part of the nearest-neighbor coupling is much greater than the imaginary part. For example, when the center-to-center distance (say, $a_x$) is 10 nm, $\mathcal{J}$ is about 0.14 meV. When the center-to-center distance is 5 nm, $\mathcal{J}$ becomes 1.1 meV.

Diagonalizing $\hat H$ yields complex eigenvalues (denoted by $\Lambda$), whose imaginary parts reflect NC decay rates. Real parts represent the corresponding energy of emitted radiation. States that decay faster or slower than a single NC are called \emph{superradiant} or \emph{subradiant}, respectively. The maximum SR rate, denoted by $\Gamma_{\text{SR}}$, is defined as  $\Gamma_\text{SR}\equiv\text{max}[-\Im{\Lambda}]/(\hbar/2)$. The corresponding, energy-equivalent, $\hbar\Gamma_{\text{SR}}$ is called the superradiant decay width. $\Gamma_{\textrm{SR}}$ is normalized by the emission rate of an individual, non-interacting NC to obtain a dimensionless SR enhancement factor $\gamma_{\textrm{enh}}=\frac{\Gamma_{\text{SR}}}{\gamma_{\text{r}}}$. We also define the spectral width as the range of the real parts of the eigenvalues, $\Delta\mathcal{E} = \text{max}[\Re\{\Lambda\}]-\text{min}[\Re\{\Lambda\}]$.

SR energy shifts are obtained from the real parts of the eigenvalues using $\Delta E = \Re{\Lambda} - E_0 $. SR redshifts of 0.3 meV are predicted, which do not reproduce the much larger redshifts observed in practice by Raino \textit{et al}., and Baranov \textit{et al}. \cite{baranov2020aging}. We have therefore previously rationalized larger than expected SR redshifts as due to SR from small, interacting NC subensembles within superlattices  \cite{mattiotti2020thermal}.

\begin{figure}[htpb]
    \includegraphics[width=0.8\linewidth]{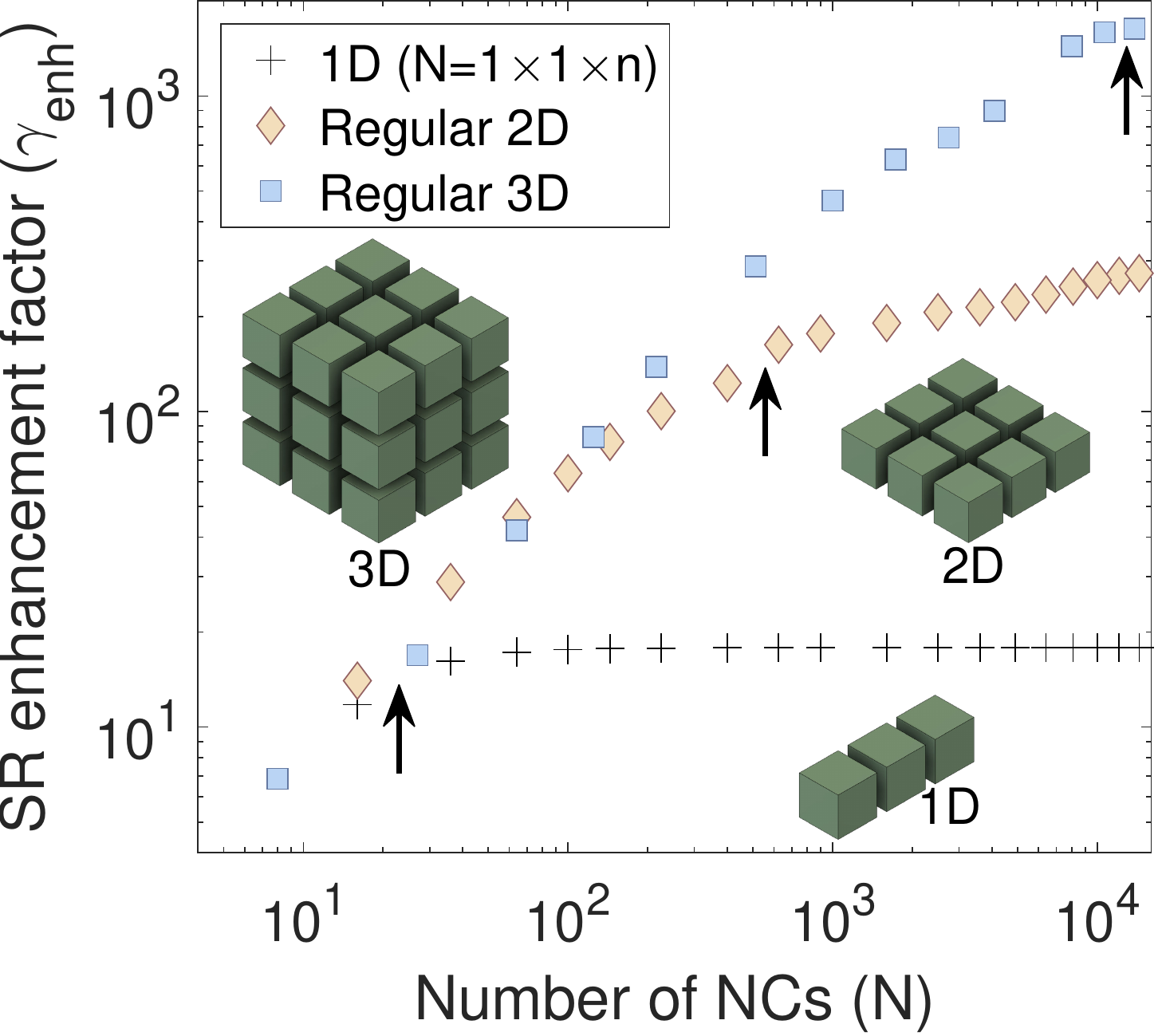}
    \caption{Comparison of the maximum possible SR enhancement from different superlattice dimensionalities composed of cube-shaped NCs with a center-to-center distance of 10 nm (i.e. $a_x=a_y=a_z=10\textrm{ nm}$). Insets show illustrations of NC superlattices with 1D, 2D and 3D dimensionalities. The arrows indicate when the length of the superlattice becomes equal to the wavelength of incident radiation ($N=23$ in 1D, $N=561$ in 2D and $N=1.33\times 10^4$ in 3D).}
    \label{fig:scaling}
    \hrulefill
\end{figure}

\section{Results and discussion}
\subsection{Superradiance in the absence of disorder}
\subsubsection{Influence of superlattice dimensionality}

\begin{figure*}[htbp]
    \includegraphics[width=0.8\linewidth]{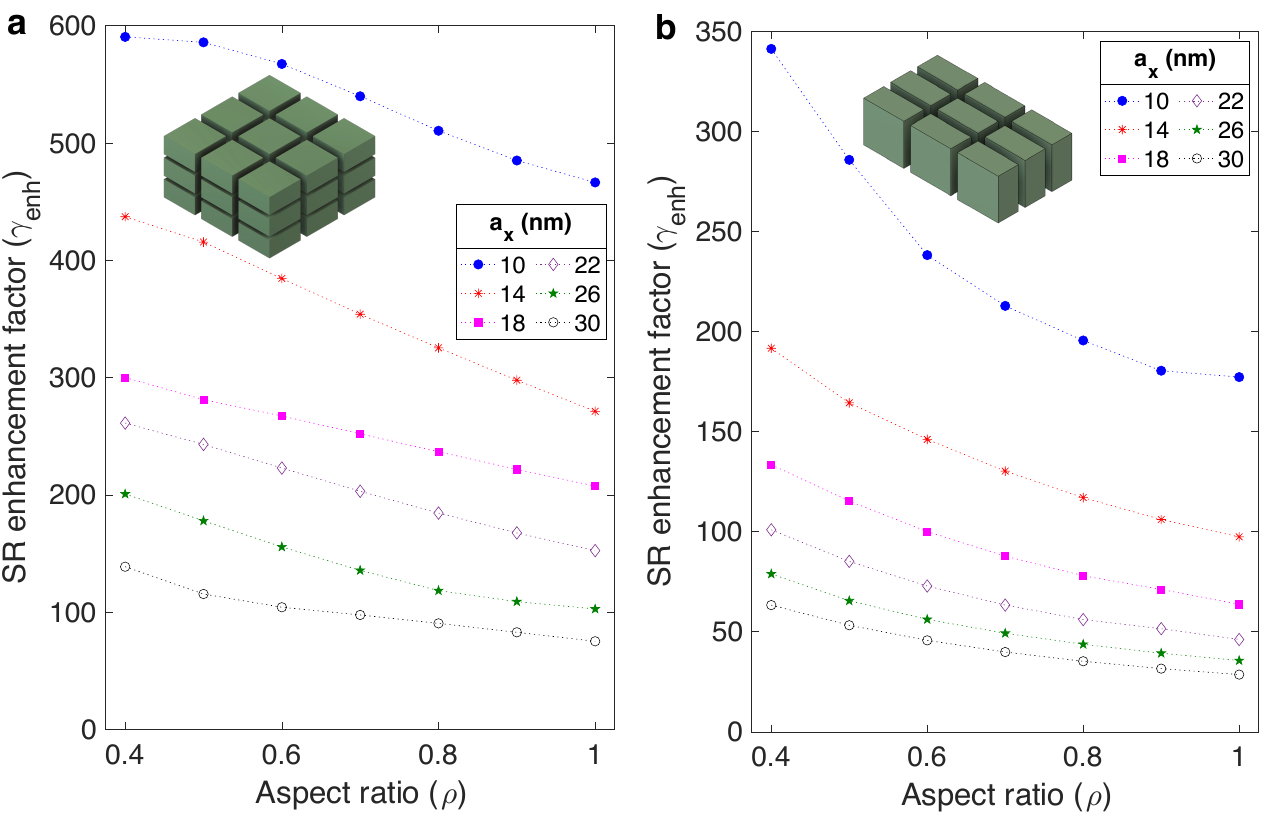}
          \caption{Comparison of the maximum possible superradiant enhancement from NCs of different shapes, all arranged in (a) a regular 3D superlattice with 10 NCs along each side ($N=1000$, $N_x=N_y=N_z=10$); (b) a regular 2D superlattice with 32 NCs along each side ($N=1024$, $N_x=N_y=32$, $N_z=1$), we use the perfect square that is closest to 1000. (Fig. \ref{fig:scaling} shows that such a small change in $N$ does not lead to any significant changes in the superradiance rate, so we can compare plots (a) and (b).) The center-to-center distance on the x-axis of the plots refers to the two longer distances ($a_x$ and $a_y$). The insets show (a) regular 3D and (b) regular 2D, superlattices composed of cuboidal NCs.}
    \label{fig:aspectratio}
    \hrulefill
\end{figure*}

\textbf{Fig. \ref{fig:scaling}} shows how superlattice dimensionality impacts $\gamma_{\textrm{enh}}$ for assemblies consisting of cube-shaped CsPbBr$_3$ NCs with a center-to-center distance of 10 nm. Superradiance from three superlattice dimensionalities is compared, where the total number of NCs ranges from 4 to $1.5 \times 10^4$. The Fig. shows that for large $N$, the 3D superlattice is the most superradiant in absolute terms. This is followed by 2D and 1D superlattices in this order. 

Our model correctly captures a deviation from the linear increase of $\gamma_{\textrm{enh}}$ \cite{mattiotti2020thermal} as the system size becomes comparable with $\lambda$ (the wavelength of radiation inside the material), see arrows in \textbf{Fig. \ref{fig:scaling}}. For the 1D superlattice, saturation occurs when $N\times a$ (the length of the superlattice) results in a physical distance of order $\lambda$. This is in agreement with an analogous result for atomic lattices \cite{sierra2022dicke}. At small $N$ where the entire superlattice is smaller than the wavelength of light, $\gamma_{\textrm{enh}}$ is roughly equal for all three superlattice dimensionalities and increases linearly with $N$.

One can also visualize the wavefunction corresponding to different eigenstates of the Hamiltonian (Eq. \ref{hamiltonian}) for 2D superlattices. Supplementary Material (SM) \cite{suppinfo} \textbf{Figs. S1 and S2} show the real and imaginary parts of the wavefunction corresponding to the most superradiant and a typical strongly subradiant state. The wavefunction corresponding to the most superradiant state has the fewest possible nodes for the size of the superlattice, while that of a subradiant state oscillates with a much smaller wavelength.

\subsubsection{Dependence on NC aspect ratio}

We now study the dependence of $\gamma_{\textrm{enh}}$ on NC aspect ratio for the 3D and 2D superlattices first considered in \textbf{Fig. \ref{fig:scaling}}. It is known that colloidal syntheses produce NCs with variable aspect ratios. Classic examples include CdSe nanorods \cite{peng2000shape}, nanoplatelets \cite{ithurria2008quasi,ithurria2011continuous}, and nanowires \cite{yu2003cadmium,liu2010origin,grebinski2004solution} and more recently, CsPbBr$_3$ platelets \cite{philbin2021room,toso2021metamorphoses}. It has also been possible to self assemble such cuboid-like NCs into superlattices \cite{philbin2021room,sui2021zone}.

In our model, $\hat H$ (Eq. \ref{hamiltonian}) depends only on the NC center-to-center distance ($\vec{r}_{mn}$). The aspect ratio ($\rho$) implicitly takes into account NC shape and inter-particle spacings due to the presence of organic ligands. In the results that follow, the number of NCs along each edge of a superlattice remains constant. Therefore, the overall aspect ratio of the superlattice is equal to the ratio of NC center-to-center distances. In all cases, NC dimensions are greater than $a_B$ so that strong confinement effects can be ignored. For the same reason, we assume that NC transition dipoles do not change significantly.

\textbf{Fig. \ref{fig:aspectratio} (a)} now reveals that as $\rho$ decreases from $\rho=1-0.4$, $\gamma_{\textrm{enh}}$ increases. The reason for this increase in $\gamma_{\textrm{enh}}$ is the decrease in NC center-to-center distance along one direction, which increases the the number of NCs (and hence emitters) per unit volume. Conversely, increasing center-to-center distances reduces NC densities, and suppresses $\gamma_{\textrm{enh}}$. \textbf{Fig. \ref{fig:aspectratio} (b)} shows that $\gamma_{\textrm{enh}}$ increases are even more significant when low aspect ratio NCs are arranged into 2D superlattices. See SM \cite{suppinfo} \textbf{Fig. S3} for the dependence of $\gamma_{\textrm{enh}}$ on $\rho$ in 1D, which is equivalent to changing the NC center-to-center distance along the superlattice direction.

Even though we predict a monotonic increase in superradiant enhancement as NC aspect ratio decreases, we have avoided more extreme aspect ratios where we expect strong confinement in one (nanosheets) or two (nanowires) dimensions \cite{shamsi2019metal,shamsi2017bright}. In such systems, the response to light of the NCs might become anisotropic.

\begin{figure*}[htpb]
    \includegraphics[width=\linewidth]{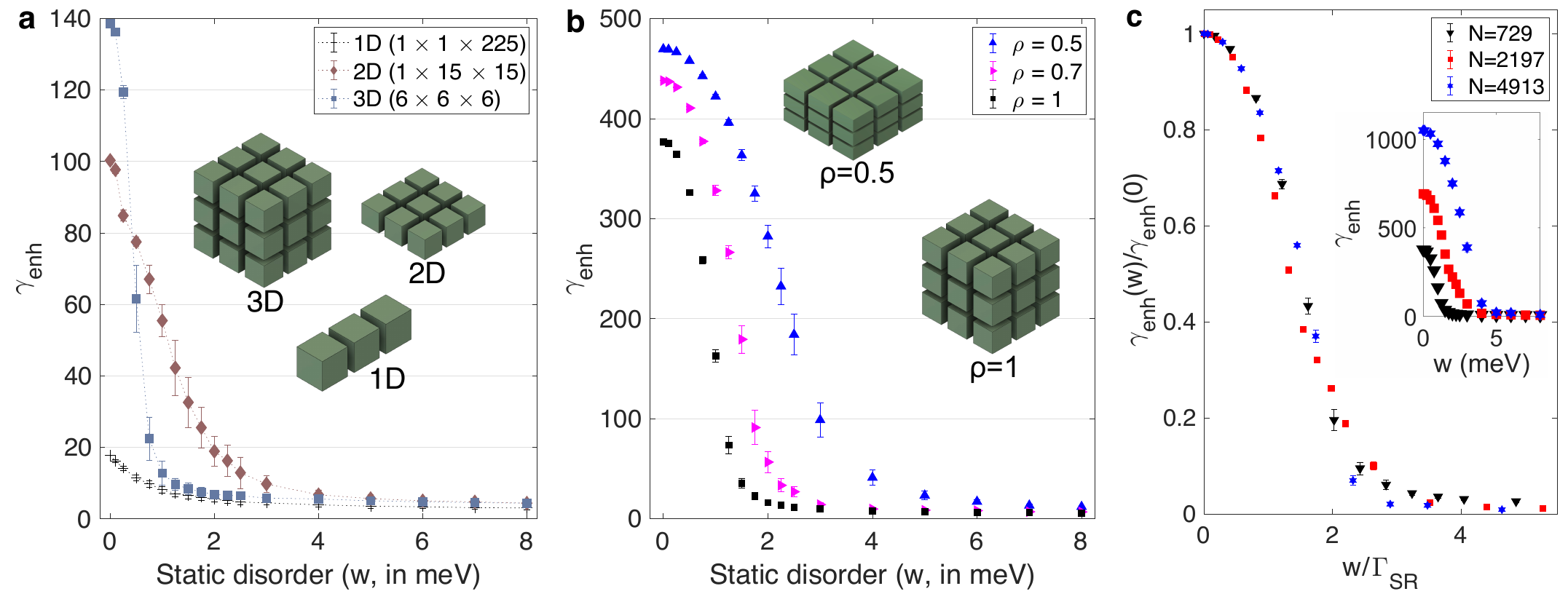}
    \caption{ Comparison of the robustness of the superradiant enhancement factor ($\gamma_{\textrm{enh}}$) to static disorder in superlattices of (a) different dimensionalities ($N=216/225$, $a_x=a_y=10$ nm), (b) different NC aspect ratios ($N=729$, $a_x=a_y=10$ nm), and (c) different $N$ (3D superlattices of cubic NCs with $a_x=a_y=a_z=10$ nm). In panel (c), the superradiant enhancement is normalized by its value in the absence of static disorder and the static disorder is normalized by the SR decay width, while the inset shows the absolute values on both axes. The error bars in all plots show the standard deviation of the eigenvalues over different disordered configurations.}
    \label{fig:std_robustness}
    \hrulefill
\end{figure*}

\subsection{Robustness to static disorder}
We now investigate SR robustness to static disorder for superlattices with different dimensionalities. In practice, a finite NC size distribution exists within a superlattice. For example, in the superlattices used by Raino \textit{et al}. \cite{raino2018superfluorescence}, the size of the NCs was 9.45 $\pm$ 0.41 nm. This size variation, in turn, introduces electronic disorder into the system due to the size-dependency of NC band gaps \cite{brennan2017origin}. In the intermediate to weak confinement regimes, this disorder is of the order of a few meV, which is much smaller than the band gap, but is comparable to the coupling between NCs.  In order to model this variation in band gaps, we add a random excitation energy $w$ to our Hamiltonian matrix (Eq. \ref{hamiltonian}). The value of $w$ ranges from 0 to 8 meV \cite{mattiotti2020thermal}. The Hamiltonian with disorder, $H_{mn}^{\alpha\beta} (w) = H_{mn}^{\alpha\beta} + \delta_{\alpha\beta}\delta_{mn}W^\alpha_m$, where $W^\alpha_m$ are uniformly distributed random real numbers on the domain $[-w/2,w/2]$. For the values of $N$ that we simulate in this paper, the superradiant enhancement becomes negligible at about 8 meV, and hence systems with even greater static disorder are not simulated.

\subsubsection{Robustness in small superlattices}

\textbf{Fig. \ref{fig:std_robustness} (a)} shows the relative robustness of 3D, 2D, and 1D cubic NC superlattices to $w$. For $N\sim 225$, a 2D superlattice shows a higher robustness to static disorder than a corresponding 3D superlattice of the same size. This robustness may be a consequence of the fact that superradiant states lie near the edges of the spectrum in 2D superlattices, where the density of states is low. On the other hand, 3D superlattices have fewer superradiant states and they lie close to the center of the spectrum, where density of states is higher. Therefore, the superradiant states in 3D superlattices are more susceptible to mixing with other eigenstates induced by static disorder. This property is illustrated in SM \cite{suppinfo} \textbf{Fig. S4}.

This increased robustness has limitations. When $N$ is large, SR from 2D superlattices becomes sub-linear and is less than that from 3D superlattices, as shown in \textbf{Fig. \ref{fig:scaling}}. A higher robustness to static disorder is insufficient to overcome this sub-linear scaling, resulting in 3D superlattices having greater superradiance despite being less robust to static disorder (see SM \cite{suppinfo} \textbf{Fig. S5}).

Next, we study the sensitivity of 3D superlattice SR robustness to NC aspect ratio and static disorder. We show results for $N=729$ because 3D superlattices are more superradiant than 2D or 1D in this regime. The results are qualitatively identical for other values of $N$ (see SM \cite{suppinfo} \textbf{Fig. S6}). \textbf{Fig. \ref{fig:std_robustness} (b)} shows that 3D superlattices, consisting of cuboidal NCs (e.g., $\rho = 0.5$), are as much as 15 times more robust than corresponding 3D superlattices made of cube-shaped ($\rho=1$) NCs for realistic \cite{mattiotti2020thermal} disorder values ($2$ meV $< w < 4 $ meV). This is to be contrasted to the $w=0$ case where only a $\sim 25\%$ enhancement is seen for $\rho=0.5$ versus $\rho=1$ [cf. \textbf{Fig. \ref{fig:aspectratio} (a)}]. This robustness to static disorder is further enhanced when cuboidal NCs are arranged into a 2D superlattice [see SM \cite{suppinfo} \textbf{Fig. S7 (a)}].

Smaller aspect ratio cuboidal NCs are more robust to static disorder due to the increased coupling between NCs. To confirm this, we plot $\gamma_{\text{enh}}$ with respect to static disorder re-scaled by the nearest neighbor coupling ($w/\mathcal{J}$). For small $N$, $\gamma_{\text{enh}}$ depends on the aspect ratio only through $w/\mathcal{J}$, as shown in SM \cite{suppinfo}\cite{suppinfo} \textbf{Fig. S6}. This simple scaling does not work for large $N$ [see \textbf{Fig. S6 (c)}]. As we discuss in the next section, cooperative robustness plays a role when $N$ is large.

\begin{figure*}[htbp]
    \includegraphics[width=\linewidth]{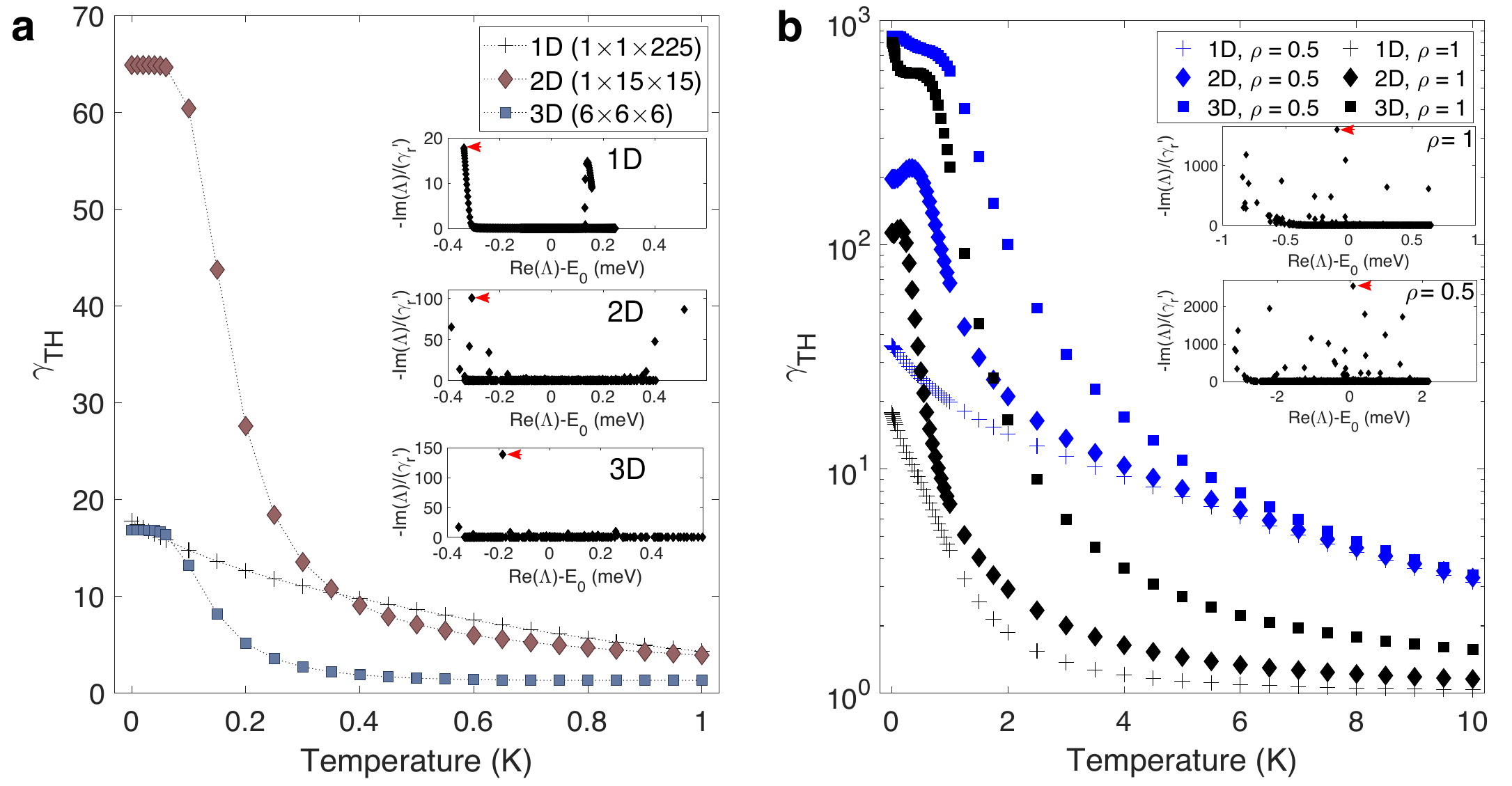}
    \caption{Comparison of the robustness of superradiant enhancement to thermal decoherence in superlattices of (a) different dimensionalities ($a_x=a_y=10$ nm, $N\simeq 225$) and (b) NC aspect ratios and dimensionalities ($a_x=a_y=10$ nm, $N \simeq 1.06\times 10^{4}$). Insets in panels (a) and (b) show the spectra of the Hamiltonian given in Eq. (\ref{hamiltonian}) for different dimensionalities and aspect ratios for the same $N$ as in the corresponding Fig.. In the insets, superradiant eigenvalues are indicated by red arrows and we have defined $\gamma_r' \equiv \hbar\gamma_r/2$.}
    \label{fig:temp_robustness}
    \hrulefill
\end{figure*}

\subsubsection{Robustness in large superlattices: cooperative robustness}

We now examine how robustness to static disorder scales with number of NCs in the superlattice for a fixed geometry. \textbf{Fig. \ref{fig:std_robustness} (c)} shows the relative robustness of SR to static disorder for different $N$ in 3D superlattices of cubic NCs. We see that as $N$ increases, robustness to static disorder $w$ [measured through $\gamma_{\text{enh}}(w)$] increases proportionally with the superradiant decay rate in the absence of disorder [$\gamma_{\text{enh}}(0)$]. This indicates the presence of cooperative robustness \cite{celardo2014cooperative,celardo2014cooperative2,chavez2019real}, which occurs when the superradiant decay width ($\hbar \Gamma_{\text{SR}}$) is much greater than the nearest-neighbor coupling ($\mathcal{J}$) \cite{celardo2014cooperative} (see SM \cite{suppinfo} \textbf{Fig. S8}). The same effect is not seen in 2D superlattices (see SM \cite{suppinfo} \textbf{Fig. S9}) due to the sublinear scaling of SR with $N$.

The origin of this effect can be qualitatively explained as follows: a very large decay width implies a large coupling of the system with the electromagnetic field. Such strong coupling protects the system from disorder, which must become comparable in magnitude to suppress SR. In more mathematical terms, one can show that a large decay rate implies that the superradiant eigenvalue is more separated from other eigenvalues of the system in the complex plane, producing a gap along the imaginary axis. When static disorder becomes comparable to the difference between the energies of the eigenvalues of the system, superradiance is suppressed. The presence of a gap along the imaginary axis, therefore protects the system with respect to disorder similar to an energy gap along the real axis \cite{chavez2019real}. 

\textbf{Fig. \ref{fig:std_robustness} (c)} shows that in 3D superlattices, $\gamma_{\text{enh}}$ is reduced to half of its disorder-free value when static disorder is about 1.5 times the superradiant decay width ($w \approx \hbar\Gamma_{\text{SR}}$). This suggests that superradiance can resist static disorder of the order of thermal energy ($k_BT$) at room temperature for a large 3D superlattice. The number of nanocrystals in realistic superlattices is at least $10^6$, for which there can be a $\gamma_{\text{enh}}$ of $10^4$ in the absence of static disorder \cite{mattiotti2020thermal}. Given that the natural decay width ($\hbar\gamma_r$) of a single perovskite nanocrystal corresponds to $1.6 \times 10^{-3}$ meV, this leads to a SR decay width ($\hbar \Gamma_{\text{SR}}$ of about $16$ meV). Thus, for $w=24$ meV (approximately the room temperature thermal energy) large superradiant enhancements are possible. For example, if the average size is 10 nm, then an NC size variation of 10\% (20\%), that is 10 nm $\pm$ 1 nm (10 nm $\pm$ 2 nm) produces a static disorder of 25 meV (50 meV) \cite{brennan2017origin}.

Cuboidal superlattices produce even stronger superradiance and as a result show an even greater cooperative robustness to static disorder (see SM \cite{suppinfo} \textbf{Fig. S9}). In large 3D superlattices of NCs with $\rho=0.5$, the SR decay width $\hbar \Gamma_{\text{SR}}$ is about $24$ meV, and they can show large SR enhancements for up to $w=48$ meV, which is approximately the static disorder when the NC size variation is 10 nm $\pm$ 2 nm. 

\subsection{Robustness to thermal disorder}

Given that any potential application of SR requires robustness to temperature, the robustness of SR to thermal disorder is investigated by defining a thermally-averaged superradiance rate \cite{mattiotti2020thermal}, denoted $\Gamma_{\textrm{TH}}$. This is an average superradiant decay rate, weighted by Boltzmann distributions in energy 
\begin{equation}\label{thermalgamma}
    \Gamma_{\textrm{TH}} = -\frac{2}{\hbar} \frac{\sum_{j=1}^{3N} \Im(\Lambda_j)e^{-\Re(\Lambda_j)/k_BT}}{\sum_{j=1}^{3N} e^{-\Re(\Lambda_j)/k_BT}}.
\end{equation}
$\Gamma_{\textrm{TH}}$ captures the effects of a non-zero temperature on SR. Equation (\ref{thermalgamma}) is only valid when thermalization, which typically occurs on the picosecond timescale \cite{brennan2017origin}, is faster than other relaxation processes in the NC. This is usually true when the superradiant decay rate is smaller than the thermalization rate ($\gamma_{\text{enh}}<10^3$) \cite{mattiotti2020thermal}.

We define a thermal enhancement factor, similar to $\gamma_{\text{enh}}$, as $\gamma_{\text{TH}}=\Gamma_{\text{TH}}/\gamma_r$. \textbf{Fig. \ref{fig:temp_robustness} (a)} now plots $\gamma_{\text{TH}}$ for different superlattice dimensionalities. We observe that 1D and 2D superlattices are more robust to thermal decoherence than 3D superlattices for $N\simeq 216$. The qualitative trend remains the same for smaller $N$. However, when $N$ becomes large (e.g., $N=729$) 3D superlattices show the greatest superradiance (see SM \cite{suppinfo} \textbf{Fig. S10}).

To understand why superradiance from 2D superlattices is more robust to thermal disorder than their 3D counterparts, we study the spectra of the corresponding Hamiltonians. The spectra of the Hamiltonians in the complex plane show the energy (real part) distribution of the superradiant decay widths (imaginary part). The energy of a superlattice's most superradiant state strongly depends on its dimensionality. 

In 1D and 2D, many superradiant states are concentrated near the edges of the eigenvalue spectrum (see insets in \textbf{Fig. \ref{fig:temp_robustness}(a)}). In 3D, however, the most superradiant state lies at the center. Eigenvalue maps in the insets of \textbf{Fig. \ref{fig:temp_robustness}(a)} show that typical spectral widths are less than 1 meV. At very low temperatures ($\ll 10$ K $= 0.86$ meV$/k_{\textrm{B}}$), the system is in one of the lowest energy eigenstates with very high probability. At room temperature or higher ($T>300$ K $= 25.8$ meV$/k_{\textrm{B}}$), the system is almost equally likely to be in any of the eigenstates. Thus, low dimensional systems are favored at low temperature and small $N$ due to the fact that there are more superradiant states close to the ground state, where the excitation concentrates. 

Next, we consider the effect of $\rho$ and $N$ on robustness to thermal disorder. \textbf{Fig. \ref{fig:temp_robustness} (b)} shows that superlattices of all dimensionalities with cuboidal ($\rho=0.5$) NCs are more superradiant than those with cube-shaped ($\rho=1$) NCs at all temperatures. This is because the lowest energy eigenvalues are more superradiant in such superlattices (\textbf{Fig. \ref{fig:temp_robustness} (b)}, inset). The spectral width in $\rho=0.5$ NC superlattices is also over two times greater than that in $\rho=1$ NC superlattices, further improving the superlattice's robustness to thermal disorder. Here, we have shown results for $N=22^3\simeq1.06\times10^4$, because 3D superlattices are more superradiant than 1D and 2D in this regime. For much smaller $N$ ($N<1000$), 3D superlattices with cuboidal ($\rho=0.5$) NCs are just as superradiant or slightly more superradiant than cubic NC superlattices at $T>2$ K (see SM \cite{suppinfo} \textbf{Fig. S11}).

However, for very low temperatures ($T\leq 2$ K), the opposite is true because lowest energy eigenvalues, which contribute the most to radiative decay rate at low temperature, are \emph{subradiant} in $\rho=0.5$ NC superlattices when $N$ is small. As we did for static disorder, we can further increase superradiance by combining the higher robustness to thermal disorder of the 2D superlattice geometry (which occurs when $N \lesssim 729$) and the higher/comparable robustness of cuboidal NC shape by arranging cuboidal NCs in a 2D superlattice (see SM \cite{suppinfo} \textbf{Fig. S7 (b)}).

The $N$-dependence of SR's robustness to thermal disorder in 2D and 3D superlattices of cubic and cuboidal NCs is shown in SM \cite{suppinfo} \textbf{Fig. S12}. $\gamma_{TH}$ increases monotonically with $N$, reaching upto $10^2$ at a temperature of about 6 K. The $\gamma_{TH}$ for large $N$ shown in \textbf{Figs. \ref{fig:temp_robustness}} and \textbf{{S12}} is only the lower bound due to cooperative robustness. 

Our assumption that thermalization is the fastest timescale fails for large system sizes where photons could be emitted superradiantly before thermalization occurs. Based on prior calculations, in the presence of static disorder of the same order of magnitude as room-temperature energy, superradiant decay widths are $\sim 16$ meV. This corresponds to a superradiant lifetime of about 80 fs, which is much shorter than typical thermalization times of $\sim 1$ ps. Studying superradiance in this regime requires a non-equilibrium simulation of systems, which will be investigated in a future study. 

\section{Conclusions and outlook}

We have analyzed the dependence of NC superlattice SR response with superlattice dimensionality, NC aspect ratio and number of interacting dipoles. We show that a dimensional crossover occurs as the number of NCs increases in the superradiance enhancement from nanocrystal superlattices. Specifically, the main results obtained are the following: 
\begin{enumerate}
    \item While one might think that 3D superlattices are always better for achieving large, robust SR responses, we show that this is not thee case for small numbers of interacting NCs. 2D superlattices made of less than 1000 NCs are more robust both to static and thermal disorder than 3D superlattices with the same number of NCs. Given that the coherent domains determining SR response in large NC superlattices can be quite small \cite{raino2018superfluorescence,mattiotti2020thermal}, our analysis can help guide future experiments.

    \item On the other hand, for large numbers of NCs, 3D superlattices always produce larger and more robust superradiance. Remarkably, we have shown that for large superlattices robustness to static disorder increases with superradiant decay width, which in 3D superlattices increases with the system size. This implies the emergence of cooperative robustness to disorder. Due to this effect we predict that large 3D superrlattices can show superradiance even at room temperature, provided that the size fluctuations of the NCs composing the superlattice can be controlled. For example, if the average size is 10 nm, then the NC size variation should be less than 10\%, that is 10 nm $\pm$ 1 nm for the static disorder to be less than 25 meV. 
    
    \item Changing NC aspect ratio can have a large impact on SR response. We show that decreasing aspect ratios by packing NCs more closely together results in stronger coupling and larger robustness of SR to both static and thermal disorder in all superlattice dimensionalities.
\end{enumerate}

Several challenges remain. If the fluence of the incident radiation is high, the single excitation assumption fails. Several NCs may be excited simultaneously, leading to a $N^2$ scaling of SR enhancement. In addition, when superradiant decay times are faster or comparable to the thermalization time, the decay rate of superlattices will deviate from the thermal average of the decay rates and the full nonequilibrium dynamics of the open quantum system must be taken into account.

\subsection*{Acknowledgements}
S.G., D.E., M.K. and B.J. thank the U.S. National Science Foundation (DMR-1952841) for financial support. We thank Fausto Borgonovi, Stefano Toso, Irina Gushchina, Zhuoming Zhang, Gabriele Rain\'{o}, Thilo St\"{o}ferle and Maksym V. Kovalenko for helpful discussions. This research was supported in part by the Notre Dame Center for Research Computing through access to key computational resources.

\end{document}